\documentclass[twocolumn]{openjournal} 

\usepackage{orcidlink} 
\usepackage[T1]{fontenc}
\usepackage{ae,aecompl}
\usepackage[T1]{fontenc}
\usepackage{ae,aecompl}
\usepackage{hyperref}
\usepackage{url}

\usepackage{longtable}
\usepackage{booktabs}
\usepackage{tabu}
\usepackage{graphicx}
\usepackage{multirow}
\usepackage{ulem}
\usepackage{color}
\usepackage{amsmath}

\def \cm{~\rm{cm}}
\def \s{~\rm{s}}
\def \km{~\rm{km}}

\def \K{~\rm{K}}

\def \g{~\rm{g}}
\def \G{~\rm{G}}

\def \erg{~\rm{erg}}

\def \yr{~\rm{yr}}

\usepackage{xcolor}
\definecolor{redak}{rgb}{0.9,0.15,0.05}

\shorttitle{Mixing NS material into jets in CEJSNe}
\shortauthors{Soker}

\graphicspath{{./}{figures/}}

\begin{document}

\title{Mixing neutron star material into the jets in the common envelope jets supernova r-process scenario}

\author{Noam Soker\,\orcidlink{0000-0003-0375-8987}} 
\affiliation{Department of Physics, Technion Israel Institute of Technology, Haifa, 3200003, Israel}

\date{\today}

\begin{abstract}
I find that the accretion disk around the neutron star (NS) that enters the core of a massive evolved star in the frame of the common-envelope jets supernova (CEJSN) r-process scenario can penetrate the crust of the NS, mix neutron-rich crust material into the disk, and enrich the jets that the disk launches with the neutron-rich material. As the NS accretes at high rates from the core inside which it revolves, it forms an accretion disk with high density. In the CEJSN r-process scenario, the very high density in the accretion disk results in low electron fraction gas, enabling the r-process. Jets carry the r-process elements out. The new claim in this study is that the high-density accretion disk destroys part of the NS crust and entrains this mass. The Kelvin–Helmholtz instability mixes material from the deeper crust. The total neutron-rich mass that the disk mixes and the jets carry can be up to $\approx 0.01 M_\odot$. Enriching the accretion disk with neutron-rich material ensures a low electron fraction as required by the r-process nucleosynthesis and the ejection of massive r-process ejecta, $0.01-0.03 M_\odot$. 
I strengthen the CEJSN r-process scenario, but do not claim it is the main r-process site.  I only claim that two or more r-process sites contribute to r-process nucleosynthesis. 
\end{abstract}

\section{Introduction} 
\label{sec:intro}

The heaviest elements in the Universe are synthesized in the rapid neutron capture process (r-process), wherein, in a neutron-rich environment, intermediate-mass elements capture tens to about two hundred neutrons in a matter of seconds. There are several theoretical r-process sites: (1) the merger of two neutron stars (NSs; e.g., \citealt{Gorielyetal2011, Wanajoetal2014, Beniaminietal2016a, Beniaminietal2016b, BeniaminiPiran2024, Jietal2016, Metzger2017, Banerjeeetal2020,  Dvorkinetal2021, vandevoortetal2022, MaozNakar2024, Qiumuetal2025}); (2) magnetorotational supernovae, i.e., core-collapse supernovae that are powered by a fixed-axis pair of jets (e.g., \citealt{Winteleretal2012, Nishimuraetal2015, HaleviMosta2018, Reichertetal2021, Reichertetal2023, Yongetal2021}); (3) winds from the newly born NS in core-collapse supernovae (e.g., \citealt{Prasannaetal2024}), possibly limited to the first r-process peak, i.e., the weak r-process (e.g., \citealt{WangBurrows2023, WangBurrows2024}); (4) common envelope jets supernovae (CEJSNe; e.g., \citealt{Papishetal2015, GrichenerSoker2019, GrichenerSoker2022RNAAS}); (5) collapsars, i.e., core-collapse supernova that forms a black hole at the center (e.g.,  \citealt{Siegeletal2019, Siegeletal2022, Braueretal2021, Issaetal2025}); (6) magnetar giant flares (e.g., \citealt{Cehulaetasl2024, Pateletal2025a, Pateletal2025b}).  

The NS-NS merger site is supported by observations that indicate the presence of very heavy radioactive elements (e.g., \citealt{Chornock_2017, kasen17, pian17, Sr19, kasliwal19, Watsonetal2019, Domoto_2022, Levanetal2024}). 
Some studies find that the NS-NS merger site cannot be the only r-process site (e.g., \citealt{Waxmanetal2018, Cote_2019, Ji_2019, Kobayashietal2020, Kobayashietal2023, holmbeck_2023}). Many studies claim the need for two or more r-process sites to explain the r-process abundances in the Galaxy (e.g., \citealt{Wehmeyeretal2015, GrichenerSoker2019paradigm, HaynesKobayashi2019, Moleroetal2021, Tsujimoto2021, Farouqietal2022, Naiduetal2022, Yamazakietal2022, Stormetal2025}). 

This study deals with the CEJSN r-process scenario (e.g., \citealt{GrichenerSoker2019}; for a review, see \citealt{Grichener2025}), which can account for many observed r-process properties (e.g., \citealt{GrichenerSoker2019paradigm}), like the europium evolution in the Galaxy \citep{GrichenerKobayashiSoker2022}, and the nucleosynthesis of the three r-process peaks \citep{JinSoker2024}. 
In the CEJSN r-process scenario, an NS enters the core of a massive star, accretes mass at a high rate, and launches jets. The jets expel some core material; the rest forms a dense accretion disk around the NS, forming neutron-rich material that enables the r-process in the disk and the base of the jets. The jets carry the r-process elements away.   
\cite{Grichener2023} finds that the rate of CEJSNe, where an NS enters a common envelope evolution with the core of a red supergiant and destroys the core to form a disk around the NS, is such that the CEJSN r-process can account for a large fraction of the r-process elements. Still, it is not the only r-process site. In the CEJSN r-process scenario, nucleosynthesis occurs in the accretion disk and the base of the jets. Past studies of the CEJSN r-process scenario considered the neutron-rich gas to result solely from the inner zone of the accretion disk, where neutrino cooling overpowers viscous heating. 
Motivated by r-process scenarios that take the neutron-rich material to be the ejected outer layers of the NS, i.e., NS wind and the magnetar giant flare scenarios, I study the possibility that the accretion disk in the CEJSN scenario entrains some material from the outer layers of the NS, enhancing r-process production, and ejecting the newly synthesized r-process elements with the jets.

\section{The accretion disk penetration to the crust} 
\label{sec:AccretionDisk}

For the density of the accretion disk and its vertical scale height, I take  equations from \cite{GrichenerSoker2019}, based on \cite{Chevalier1996}, 
but scale the radius to the surface of the NS at $r=12 \km$, 
 \begin{equation}
\begin{split}
\rho_{\rm d} \approx 10^{12} &   
\left( \frac{\dot M_{\rm acc}}{0.05 M_\odot \s^{-1}} \right)^{0.84}
\left(\frac{M_{\rm NS}}{1.4 M_\odot} \right)^{0.76}
\\
&
\times
\left(\frac{r}{12 \km} \right)^{-2.29}
\left(\frac{\alpha}{0.01} \right)^{-1} \g \cm^{-3} ,
\end{split}
\label{eq:density}
\end{equation}
and
\begin{equation}
\begin{split}
\frac{H}{r} \approx 0.29
\left( \frac{\dot M_{\rm acc}}{0.05 M_\odot \s^{-1}} \right)^{0.05} &
\left(\frac{M_{\rm NS}}{1.4 M_\odot} \right)^{-0.42}
\\
&
\times
\left(\frac{r}{12 \km} \right)^{0.26}.
\end{split}
\label{eq:Thickness}
\end{equation}
where $M_{\rm NS}$ is the NS mass, $\alpha$ is the disk viscosity parameter, and $\dot M_{\rm acc}$ is the mass accretion rate by the NS   scaled by the typical value for the present scenario as \cite{GrichenerSoker2019} give.   

 I check the possibility that a strong magnetic field truncates the accretion disk. Using the truncation radius expression (e.g., equation 1 in \citealt{Longetal2005}), for a surface NS magnetic field $B_{\rm NS}$, I find the truncation radius to be 
 \begin{equation}
\begin{split}
\frac{R_t}{R_{\rm NS}} \simeq 0.5 
\left( \frac{\dot M_{\rm acc}}{0.05 M_\odot \s^{-1}} \right)^{-2/7}
\left(\frac{B_{\rm NS}}{10^{15} \G} \right)^{4/7} 
\end{split}
\label{eq:Rt}
\end{equation}
where I substituted $M_{\rm NS}=1.4 M_\odot$ and for the NS radius $R_{\rm NS}=12 \km$. I find that only if the NS magnetic field is extremely large can it truncate the accretion disk outside the NS. I consider this possibility unlikely.

In this preliminary study of the mixing process, I neglect relativistic effects. The last stable orbit of a non-rotating NS of $M_{\rm NS}=1.4 M_\odot$ is $R_{\rm s} = 12.4 \km$.
However, since the NS in the CEJSN scenario already accreted mass from the envelope of the red supergiant, it is expected to rotate. A prograde orbit has a smaller last stable orbit. To have the last stable orbit for a prograde accretion disk at $R_{\rm NS}=12 \km$ I calculate the rotation velocity of the NS to be $v_{\rm cr, s} = 0.07 (G M_{\rm NS} /R_{\rm NS})^{1/2}$, for an NS with $M_{\rm NS}=1.4 M_\odot$ and a moment of inertia of $1.4 \times 10^{45} \g \cm^2$ (e.g. \citealt{Worleyetal2008}). Namely, the NS should rotate at more than about $7\%$ of its break-up velocity. However, even if the NS rotates more slowly and the last stable orbit is located outside the NS surface, the material will still be accreted, forming a rotating equatorial belt above the NS. A strong shear will occur between the accretion belt and the NS crust. 
 The magnitude of relativistic effects is $\simeq 2GM_{\rm MN}/(c^2 R_{\rm NS}) = 0.34$ for $M_{\rm NS}=1.4 M_\odot$ and $R_{\rm NS} = 12 \km$. This implies that the non-relativistic treatment presented here is accurate to tens of percent.  
To the accuracy of the present approximate study, I neglect relativistic effects. Numerical simulations of the mixing process I study here should include them.

The high density is significant for the r-process in the CEJSN r-process scenario \citep{GrichenerSoker2019}. I find that the very high density inside the accretion disk on the NS surface has another implication. In a model of a cold NS (e.g., \citealt{ChamelHaensel2008Review}), the region with densities $\rho_{\rm NS} \lesssim 10^{11} \g \cm^{-3}$, but residing below the NS envelope, is the outer crust and is dominated by neutron-rich nuclei. The region with densities $10^{11} \g \cm^{-3} \lesssim \rho_{\rm NS} \lesssim 10^{14} \g \cm^{-3}$ is the inner crust and is dominated by neutrons.
The accretion disk can penetrate the depth where the NS and disk densities are about equal. For the typical density of the CEJSN r-process scenario, as equation (\ref{eq:density}) gives, the accretion disk can penetrate the inner crust and entrain neutron-rich material.

For the pressure in the zones near the transition from the inner to outer crust, I fit a line to the profile from figure 29 of \cite{ChamelHaensel2008Review} as 
\begin{equation}
P_{\rm cr} \simeq 1.5 \times 10^{30}     
\left( \frac{\rho_{\rm cr}}{10^{12} \g \cm^{-3}} \right) 
\erg \cm^{-3}. 
 \label{eq:Pcr}
\end{equation}
The sound speed in this region is $C_{\rm s,cr} =(\gamma P_{\rm cr} /  \rho_{\rm cr}  )^{1/2} \simeq 10^4 \km \s^{-1}$. 
For the mass above radius $r$, I adopt an equation from 
\cite{Cehulaetasl2024} 
\begin{equation}
M_{\rm cr} (>r) \approx  \frac{4 \pi R^4_{\rm NS}}{G M_{\rm NS}} P_{\rm cr}(r). 
 \label{eq:Mcr1}
\end{equation}
I substitute relation (\ref{eq:Pcr}) and take an NS with mass and radius of $M_{\rm NS}=1.4 M_\odot$ and $R_{\rm NS}=12 \km$, respectively, and find 
\begin{equation}
M_{\rm cr} (>\rho) \approx 
 10^{-4}      
\left( \frac{\rho_{\rm cr}}{10^{12} \g \cm^{-3}} \right) 
M_\odot. 
 \label{eq:Mcr2}
\end{equation}

My claim here is that the accretion disk penetrates to depth $\rho_{\rm cr} \simeq 10^{12} \g \cm^{-3}$, and the jets from the inner disk remove this mass. Note that this is a very thin layer on the NS surface that the disk destroys, $\simeq 0.1 \km$. 
For the typical mass accretion rate during the r-process event,  $\dot M_{\rm acc} \simeq 0.05 M_\odot \s^{-1}$, the accretion gas replenishes this mass on a timescale of $\tau_{\rm rep} (12) \simeq M_{\rm cr}(12)/\dot M_{\rm acc} \simeq 0.002 \s$, where the number 12 inside the parenthesis refers to zones with densities below $10^{12} \g \cm^{-3}$. The sound crossing time from the accretion disk to the poles is $\simeq \pi R_{\rm NS} / 2 /C_{\rm s,cr} \simeq 0.002 \sec$. This crust zone has time to continuously rebuild itself during the accretion process that lasts for $\approx 10-100 \s$.

\section{Mixing neutron star material into the disk} 
\label{sec:KHI}

Mixing crust material penetrates deeper into the NS  because of the Kelvin-Helmholtz instability (KHI);  for an experimental demonstration, see, e.g., \cite{StrangFernando2001}, and for numerical simulation of the KHI mixing in an accreting white dwarf see, e.g., \cite{Bellomoetal2024}.  
I turn to calculate the depth of the mixing by the KHI. 

The disk rotates at the velocity $v_{\rm d}$ and the crust in the mixing layer at $v_{\rm cr} < v_{\rm d}$. 
Neglecting the effects of magnetic fields, KHI occurs for short wavelengths, $\lambda = 2 \pi /k$, that is given by the usual KHI criterion without tension 
\begin{equation}
  \lambda < \frac{ 2 \pi \rho_{\rm d} \rho_{\rm cr} 
    ( v_{\rm d} - v_{\rm cr})^2 } { g ( \rho_{\rm cr} - \rho_{\rm d}) ( \rho_{\rm cr} + \rho_{\rm d}) }.
    \label{eq:KHI0}
\end{equation}
Substituting $g=G M_{\rm NS}/R^2_{\rm NS}$ and scaling the velocity difference by the Keplerian velocity, which is also the approximate accretion disk velocity, 
\begin{equation}
\Delta q_v \equiv \frac {v_{\rm d} - v_{\rm cr} }{ \sqrt{G M_{\rm NS}/R_{\rm NS}}} \simeq 1- \frac {v_{\rm cr} }{ \sqrt{G M_{\rm NS}/R_{\rm NS}}} ,
\label{eq:DeltaqV}
\end{equation}
equation (\ref{eq:KHI0}) becomes 
\begin{equation}
  \frac{\lambda}{R_{\rm NS}} < \frac{ 2 \pi q_\rho 
    (\Delta q_v)^2 } {( q_\rho - 1) ( q_\rho + 1 ) }, 
    \label{eq:KHI1}
\end{equation}
where 
\begin{equation}
 q_\rho \equiv \rho_{\rm cr} (r) / \rho_{\rm d}. 
    \label{eq:qRho}
\end{equation}
I recall that I neglect relativistic effects (Section \ref{sec:AccretionDisk}), and take for the rotation velocity of the accretion disk on the surface of the NS $v_{\rm d} \simeq \sqrt{G M_{\rm NS}/R_{\rm NS}}$. 

To estimate the KHI's mixing depth, I take it to be of the order of the wavelength (although this does not need to be the case). The mixing depth is small compared with $R_{\rm NS}$, and I calculate near the surface of the NS. Substituting $dM_{\rm cr} =4  \pi R^2_{\rm NS} \rho_{\rm cr} (r) dr $ in the derivative $dM_{\rm cr} (> \rho) / d \rho_{\rm cr}$ by equation (\ref{eq:Mcr2}), gives    
\begin{equation}
\frac{d \rho_{\rm cr}}{\rho_{\rm cr}}  \approx 4 \pi  R^3_{\rm NS}  \frac {10^{12} \g \cm^{-3}}{10^{-4} M_\odot} \frac {dr}{R_{\rm NS}} .  
    \label{eq:dRhoDR}
\end{equation}
Substituting $R_{\rm NS} = 12 \km$, taking $dr = \lambda$ according to the assumption, and approximating the value of $d \rho_{\rm cr}$ as the difference between the density at the mixing depth and the disk density, equation (\ref{eq:dRhoDR}) becomes 
\begin{equation}
\frac{d \rho_{\rm cr}}{\rho_{\rm cr}} \simeq 
   \frac{\rho_{\rm cr} - \rho_{\rm d}} {(\rho_{\rm cr} + \rho_{\rm d})/2 }
\approx 109.2 \frac {\lambda}{R_{\rm NS}} .  
    \label{eq:dRhoDR2}
\end{equation}
 Multiplying equation (\ref{eq:KHI1}) by 109.2 and substituting for its left-hand side from equation (\ref{eq:dRhoDR2}) with the definition (\ref{eq:qRho}) gives the condition for instability on the density at the mixing depth 
\begin{equation}
\frac{q_\rho - 1 }{ q_\rho + 1 } \lessapprox 343 \frac{ q_\rho 
    (\Delta q_v)^2 } {( q_\rho - 1) ( q_\rho + 1 ) }, 
    \label{eq:Condition1}
\end{equation}
which is simplified to 
\begin{equation}
\frac{ (q_\rho - 1)^2 }{ q_\rho} \lesssim 343 (\Delta q_v)^2 . 
    \label{eq:Condition2}
\end{equation}
I expect the NS to be rapidly rotating but not close to its break-up velocity because the NS is old and has had time to lose angular momentum, and the jets carry away a large fraction of the angular momentum in the accretion disk.  If the NS is close to its break-up velocity, its oblate structure should be considered. For $\Delta q_v \lesssim 0.8$, the solution gives large values of $q_\rho > 15$, and for the accuracy of this approximate treatment, I can further approximate equation (\ref{eq:Condition2}) to obtain a simple condition for the KHI   
\begin{equation}
q_\rho \lessapprox 86 \left( \frac{ \Delta q_v}{0.5} \right)^2 . 
    \label{eq:Condition3}
\end{equation}
For $q_{\rho} =15-86$, equation (\ref{eq:dRhoDR2}) gives $\lambda/R_{\rm NS} \approx 0.016-0.018$. By equation (\ref{eq:dRhoDR2}), the density scale height is $H_\rho \approx 0.009R_{\rm NS}$. A more accurate treatment should yield a deeper mixing length for these values of $q_\rho$, $\lambda \approx (2.5-4.5) H_\rho \simeq (0.025- 0.04 R_{\rm NS})$.

Magnetic fields can stabilize the KHI (depending on the direction of the magnetic field). To play a significant stabilizing role the condition is that the Alfven speed in the disk $v_{\rm A}=B/\sqrt{4 \pi \rho_d}$ will not be much smaller than the shear velocity $v_d -v_{\rm cr}$; the magnetic field can suppress the instability if $v_{\rm A} \ge  (v_d -v_{\rm cr})/\sqrt{2}$ (e.g., \citealt{Prajapatietal2009}). 
For the parameters used here, $\rho_{\rm d}=10^{12} \g \cm^{-3}$, $v_{\rm A}= 2820 (B/10^{15} \G) \km \s^{-1}$, and $v_{\rm Kep}= 1.24 \times 10^5 \km \s^{-1}$, the condition on the magnetic field to substantially reduce the instability is 
 \begin{equation}
B \gtrsim  3 \times 10^{16}  \Delta q_v   \G . 
    \label{eq:BKHI}
\end{equation}
This value is much larger than typical magnetic fields in NSs, even of magnetars and even for shear velocity as small as $\Delta q_v \simeq 0.3$. The magnetic field is unlikely to change much the instability unless the NS rotates close to its break-up velocity. I consider this unlikely.  
 
The conclusion from the above derivation is that the KHI might mix NS material down to $\rho_{\rm cr,mix} \approx {\rm few} \times 10 \rho_{\rm d} \approx {\rm few} \times 10^{13} \g \cm^{-3}$. Substituting $\rho_{\rm cr}=\rho_{\rm d} q_\rho$ from equation  (\ref{eq:Condition3}) with $\rho_{\rm d}=10^{12} \g \cm^{-3}$ in equation (\ref{eq:Mcr2}), I find that the entrained mass from the cool original material of the NS can be as large as  
\begin{equation}
M_{\rm cr, mix,0} \approx q_\rho 10^{-4} M_\odot \approx 0.01 \left( \frac{ \Delta q_v}{0.5} \right)^2  M_\odot  .  
\label{eq:Mmix}
\end{equation} 

In the CEJSN r-process scenario, the launching of jets that carry away r-process elements lasts for several times the dynamical time of the core of the RSG star, about several seconds (e.g., \citealt{GrichenerSoker2019}). This is much longer than the dynamical time of the NS, and the accreted material can replenish the original entrained NS layers. The newly accreted material from the accretion disk mixes and entrains NS material that the NS has just accreted. Overall, the total entrained high-density material might be $> 0.01 M_\odot$, possibly as large as $M_{\rm cr, entrain} \approx  0.03 M_\odot$. 

 The present study does not go into the details of the mixing process of the disk material with the high-density NS material, which is a solid. At densities of $\rho_{\rm cr,mix} \approx {\rm few} \times 10^{13} \g \cm^{-3}$ the solid-liquid phase transition occurs at a temperature of $T_{\rm SL} \lesssim 10^{10} \K$ (e.g., \citealt{PerezAzorinetal2006, Carreauetal2020}). A simple calculation that does not include neutrino cooling shows that the dissipation of the kinetic energy of the disk material at a density of $10^{12} \g \cm^{-2}$ heats the gas to a temperature of $T>10^{11} \K$. However, neutrino cooling reduces this temperature. Nonetheless, the dissipated energy can turn the solid mixed crust material to liquid.  

 The accretion rates of the studied scenario are highly super-Eddington. \cite{Fujimoto1993} derived an expression for the NS density at the mixing depth for much lower accretion rates (equation 36 there); the expression normalized to the Eddington mass accretion rate in that paper reads
\begin{equation}
\begin{split}
\rho_{\rm F93} \simeq &  3 \times 10^3 \alpha ^{-1} 
\left(\frac{\Omega}{\Omega_{\rm K}} \right)^{-1}  
\\ \times &
\left( \frac{\dot M_{\rm acc}}{5 \times 10^{-16} M_\odot \s^{-1}} \right)
\g \cm^{-3},
\end{split}
\label{eq:RhoF93}
\end{equation}
where $\alpha <1$ is the turbulent viscosity coefficient, $\Omega$ the NS spin angular velocity, and $\Omega_{\rm K}$ is the Keplerian angular velocity at the NS surface. 
For the accretion rate of the present scenario $\dot M_{\rm acc} \approx 0.05 M_\odot \s^{-1}$, this expression gives a density of $ \rho_{\rm F93} > 3 \times 10^{17} \g \cm^{-3}$ because $\alpha<1$ and $\Omega < \Omega_{\rm K}$. This NS density at the mixing depth is much larger than at the NS center. This expression does not hold for high accretion rates of the present scenario; however, it does suggest that the density at the mixing depth I derive here,  $\rho_{\rm cr,mix} \approx {\rm few} \times 10^{13} \g \cm^{-3}$, is reasonable, and might even underestimate the penetration depth.

\section{Implications to the r-process nucleosynthesis} 
\label{sec:Rprocess}

In the CEJSN r-process scenario, the NS enters the core of a massive star, accretes mass at a high rate, and launches jets. The jets expel some of the core material; the rest forms an accretion disk of 
$M_{\rm disk} \approx 1 M_\odot$ around the NS and launches two opposite jets carrying a total mass of $M_{\rm 2j} \approx 0.1 M_\odot$ (e.g., \citealt{GrichenerSoker2019}), which is the nominal fraction of mass that jets remove from accretion disks. The high mass accretion rate $\dot M_{\rm acc} \approx 0.05 M_\odot \yr^{-1}$  lasts $\approx 10 - 100 \s$. The NS might accrete enough mass to collapse into a black hole. During the high mass accretion rate, nucleosynthesis occurs. 

\citealt{GrichenerSoker2019} scale the density and temperature in the accretion disk at a radius of $r=50 \km$. The conditions of the accretion disk at this radius can lead to efficient r-process nucleosynthesis of the entire range of the r-process, including the third peak and tracers of the actinides \citep{JinSoker2024}. In this study, I suggest that, in addition, the accretion jet launches jets from the boundary layer between the accretion disk's inner zone and the NS's surface. The total material that the KHI and turbulence mix to the accretion disk is non-negligible relative to the total disk mass. If the accretion disk efficiently launches jets from the boundary layer, the mixed material can be a substantial fraction of the jets' material. Taking a value between the electron fraction in the crust and in the disk as \cite{GrichenerSoker2019} gives $Y_e \equiv N_e/(N_p+N_n) \simeq 0.1-0.3$; here $N_e$, $N_p$, and $N_n$ are the electron, proton, and neutron density, respectively. Therefore, the entrained material is neutron-rich, as required for the r-process; 

The entrainment of a substantial mass of neutron-rich material from the NS has two positive implications for the CEJSN r-process scenario: (1) it ensures low values of $Y_e$, and (2) it increases the amount of neutron-rich material in the disk and the jets, the material that the r-process requires. The expected r-process element mass that one CEJSN r-process scenario forms, based on crude estimates, is $M_{\rm rp,CE} \approx 0.01-0.03 M_\odot$ \citep{GrichenerSoker2019, GrichenerSoker2019paradigm}. 

 The exact calculation of the electron fraction of the mixed material that the jets carry and the final mass of the r-process elements that the jets carry out from the star is a subject for future studies. Here, I only note that the neutron fraction in the scenario I study is sufficiently high to synthesize r-process elements, as discussed by \cite{GrichenerSoker2019}.   
\cite{Winteleretal2012} find that when $Y_e  \lesssim 0.15$ inside the jets that a NS in MHD-driven supernova launches, the jets synthesize the second and third peaks of the r-process (i.e., heavy r-process). 
\cite{GrichenerSoker2019} found, based on \cite{Kohrietal2005}, that an accretion rate of $\dot M_{\rm NS} \gtrsim 0.06 M_\odot \s^{-1}$ yields $Y_e \lesssim 0.15$. 
In a more recent study, \cite{Siegeletal2019} find that an even lower mass accretion rate can yield the strong r-process, $\dot M_{\rm NS } \gtrsim 0.002 M_\odot \s^{-1}$. In all the cases that \cite{GrichenerSoker2019} considered, the accretion rate is above 10 times this lower mass accretion rate limit. The mixing I study here takes place in the innermost zone of the accretion disk. The accretion disk launches material from somewhat larger radii as well. \cite{GrichenerSoker2019} find that the value of $Y_e$ is sufficiently small out to a disk radius of $r \simeq 100 \km$; they consider the accretion disk to launch jets in the region $r \simeq 12-100 \km$. 
My point is that the accretion disk already had a low value of $Y_e$ for the r-process before I considered mixing. As emphasized above, the mixing of NS crust material ensures the low value of $Y_e$ even for somewhat lower accretion rates than the threshold above, and increases the r-process mass yield.

 The proposed scenario assumes that the crust-disk mixed material is ejected from the disk and is part of the jets' material. The motivation for this assumption is the extremely large shear in the mixing layer, which is the boundary layer between the star and the accretion disk. This boundary layer dissipates kinetic energy into thermal and magnetic energies, crucial ingredients in launching powerful jets. Outer zones of the disk also launch material into the jets. Still, the mixed material might have the lowest value of $Y_e$ and highest initial densities, and be responsible for the nucleosynthesis of the heaviest r-process elements. These are topics for future studies.

The entrainment process ensures a large mass of r-process elements per event. Observations of old stars with very low iron abundance exhibit a large scatter, suggesting that early r-process events were rare (e.g., \citealt{Francoisetal2007, BeniaminiHotokezaka2020}), particularly at the sites of the third peak (strong r-process). The rare occurrence of r-process events requires, in turn, that each r-process event forms a relatively large mass of r-process elements. 

The crust material that the accretion disk entrains in the CEJSN scenario is the same material that the recently proposed magnetar-flare r-process scenario ejects by magnetic activity (e.g., \citealt{Cehulaetasl2024, Pateletal2025a, Pateletal2025b}). However, the magnetar-flare scenario, which involves no mass accretion, yields an r-process element mass of only $10^{-5}-10^{-3} M_\odot$ over the lifetime of the magnetar's powerful flares \citep{Pateletal2025a}. This low mass implies that if it accounts for a substantial fraction of the r-process, the magnetar flare r-process site cannot be as rare as scenarios that yield much larger r-process mass per object, i.e., $\gtrsim 0.01M_\odot$, e.g., the collapsar, the NS-NS merger, and the CEJSN scenarios. Due to the large event rate, it is unclear whether the magnetar flare scenario can account for the large scatter of the r-process to iron ratio in low-metallicity stars.  

The magnetar flare scenario might encounter another challenge. According to \cite{Pateletal2025a}, the magnetar experiences $\simeq 10-1000$ giant flares over its lifetime of $\simeq 10^4 \yr$, following the formation of the magnetar in a core-collapse supernova explosion; the flares eject the r-process elements at velocities of $\approx 0.1 c$. The r-process elements catch up with the supernova ejecta and mix with it, including the iron that the supernova explosion produces. Some studies (e.g., \citealt{qianWasserburg2007, MaciasRamirezRuiz2019}) claim that the r-process sites, at least in the early Universe, are spatially uncorrelated with sites of iron production. The NS-NS merger and CEJSN scenarios are compatible with this claim, as they produce negligible amounts of iron. The magnetorotational supernovae and winds from the newly born NS scenarios occurring within a core-collapse supernova are incompatible with this claim; \cite{MaciasRamirezRuiz2019} claim the same for the collapsar scenario. I claim the same for the magnetar flare scenario here. Although NS-NS mergers and accretion-induced collapse of a white dwarf can also form magnetars (e.g., \citealt{Margalitetal2019}), these processes have long delays after star formation and are inefficient in the young Universe. 

\section{Discussion and Summary} 
\label{sec:Summary}

I examined the process of mixing neutron-rich NS crust material to the inner zone of the accretion disk in the CEJSN r-process scenario. For the typical expected accretion disk properties (Section \ref{sec:AccretionDisk}), I found that the accretion disk destroys the outer crust down to about the density of the disk. I also found that KHI can mix material deeper into the NS to densities that are tens of times higher than the accretion disk density (equation \ref{eq:Condition3}). The mixed mass from the original cold NS can be as high as $M_{\rm cr, mix,0} \approx 0.01 M_\odot$ (equation \ref{eq:Mmix}). The accretion process replenishes the crust zones the jets remove, and the mixing continues. Therefore, the total entrained material that the jets carry might, by a crude estimate, be as high as  $M_{\rm cr, entrain} \approx  0.03 M_\odot$ (conclusion of Section \ref{sec:KHI}).  

Overall, mixing the NS material with the inner accretion disk material and ejecting the mixture in jets (entrainment) makes the conditions for r-process in the CEJSN r-process scenario more favorable and increases the possible r-process element mass per event.  

Full numerical simulations of the process I studied here, and the CEJSN r-process scenario in general, should have the resolution to resolve the mixing layer and the density scale height of $H_\rho \approx 0.01R_{\rm NS}$. They should also include the accretion disk around the NS, namely, a numerical grid size of $> R_{\rm NS}$. The dynamical range is large, three orders of magnitude, implying that the CEJSN r-process scenario is difficult to simulate. 

The results of this study strengthen earlier claims (e.g., \citealt{GrichenerSoker2019paradigm}) that the CEJSN r-process contributes to r-process nucleosynthesis, particularly in the young Universe. Because of its short time delay from star formation to the r-process event, studies of the r-process in the early Universe and young Milky Way galaxy should not ignore the CEJSN scenario. However, I cannot claim that the CEJSN is the main r-process site. I only claim that two or more r-process sites most likely exist. The community should be open to and consider the six scenarios I listed in Section \ref{sec:intro} and more if they exist and I missed them. The different scenarios might lead to different abundance ratios of r-process elements, e.g., \cite{JinSoker2024} who compared four r-process sites (NS-NS merger; magnetorotational supernovae; collapsar; CEJSN) and found differences in the iridium to europium ratio and different ratios between the lanthanide mass and the total mass of r-process elements. \cite{JinSoker2024} find that a single r-process scenario from these four cannot account for all the observed stars; one must consider two or more r-process scenarios.   
 
 \acknowledgments
I thank Aldana Grichener for her very useful comments and discussions,  and two anonymous referees for their valuable comments that improved the presentation of the arguments. A grant from the Asher Space Research Institute at the Technion supported this study. 


\end{document}